# A Harmonic-based Fault detection algorithm for Microgrids

Wael Al Hanaineh
*Electric Engineering Department*
*Polytechnic University of Catalonia*
Barcelona, Spain
wael.hasan.ahmad.al.hanaineh@upc.edu

Jose Matas
*Electric Engineering Department*
*Polytechnic University of Catalonia*
Barcelona, Spain
jose.matas@upc.edu

Jorge. Elmariachet
*Electric Engineering Department*
*Polytechnic University of Catalonia*
Barcelona, Spain
jorge.el.mariachet@upc.edu

Josep.M. Guerrero
*Department of Energy Technology*
*Aalborg University*
Aalborg, Denmark
joz@et.aau.dk

*Abstract*— The trend toward Microgrids (MGs) is significantly increasing by employing Distributed Generators (DGs) which leads to new challenges, especially in the fault detection. This paper proposes an algorithm based on the Total Harmonic Distortion (THD) of the grid voltages to detect the events of faults in MGs. The algorithm uses the THD together with the estimate amplitude voltages and the zero-sequence component for the detection and identification of the faults. The performance is evaluated by using MATLAB/Simulink simulations to validate the capability for detecting different fault types in the least possible time.

## I. INTRODUCTION

A microgrid (MG) is a low-voltage power network with some distributed generations (DGs) and a cluster of loads that can run in parallel with the utility grid or independently. MGs are a means to increase the distributed penetration of renewable energy, such as photovoltaic and wind systems into the electrical grid. MGs have a positive impact on the environment and the economy by supplying power locally, eliminating losses in the lines, and offering continuous energy supply with improved reliability and efficiency [1].

MGs have several technical challenges related to its operational modes and characteristics, being the protection system a major issue, since should be protected from different kind of faults [2]. The magnitude and direction of fault currents in a microgrid change depending on the system configuration, due to the bi-directional power flow from the loads and the generators passing through the protective devices (PDs) [3]. The grid is the source of the majority of faults during grid-connected mode of operation, which results in very large fault currents. However, for islanding mode operation, the faulted currents are much smaller, due to the power limitations of semiconductor devices, which could be not enough to trigger a breaker [4]. Under such circumstances, the traditional PDs are unable to acquire enough information about the fault to detect the problems caused, which could lead to equipment instability and damage [5].

In recent years, several fault detection methods have been proposed for protecting microgrids, which can be grouped into differential, voltage, adaptive, and harmonics methods, each of them having advantages and weak points. Differential based methods are relatively simple [6], giving a high speed and sensitive fault detection response, and being unaffected by changes in the current's flow direction and magnitude. However, they have problems due to the rely on communication channels and because of imbalances and transients. Voltage based methods [7] have a good ability for preventing blackout in the system. But, they have inadequate sensitivity in grid-connected mode, and the voltage drops induced by faults can create errors. Adaptive based methods [8-10] are those in which the relay settings are automatically readjusted to be compatible with the power system conditions. However, they require a fault analysis and computation for determining the relay settings, as well as prior knowledge for network upgrades. Harmonic based methods [11-13] use the harmonic content of voltage and current for the fault protection. In [11], the Fourier transform (FFT) is used for achieving the required harmonics and, therefore, compute the THD and define the protection algorithm. However, the FFT implementation supposes a high computational burden for a digital processor. In [12], a cost-effective solution is introduced for microgrid protection using a new relay to detect and isolate the fault by injecting harmonic signals. Therefore, it acts like a directional relay with no need for any voltage transformer. Another interesting approach is presented in [13], which involves introducing a certain amount of a fifth harmonic to the fault current so that the protection device can identify the fault based on the low harmonics extracted using a digital relay with a FFT.

This paper presents a microgrid fault detection algorithm based on the measured THD levels of the grid voltages, the estimated amplitude voltages, and the zero-sequence components to detect, and identify, the faults that could happen in different locations of a MG [14]. This paper is submitted as a part of the Ph.D. of Electrical Engineering work at Polytechnic University of Catalonia. The rest of the paper is structured as follows: Section II presents the proposed detection algorithm in detail. Section III presents a MG as a case study to test the approach. Simulations are carried out to validate the performance of the proposed algorithm in section IV. Finally, the conclusion is provided in section V.

## II. FAULT DETECTION ALGORITHM

Faults might occur in one or more phases of the grid to the ground, or between phases only. Then, in this paper, an algorithm is defined using three stages for fault detection. Fig. 1 depicts the block diagram of the algorithm.

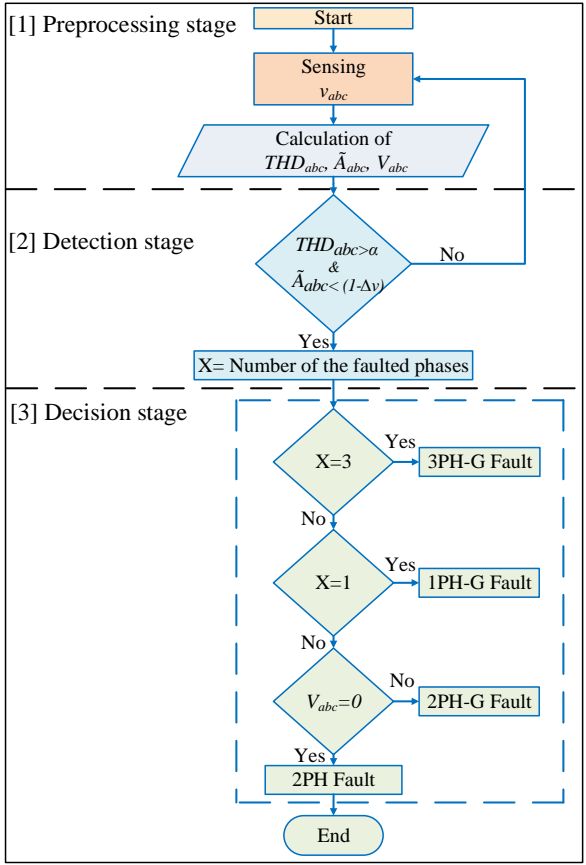

Fig. 1. Flowchart of the fault detection algorithm

## A. Pre-processing stage

In this stage, see Fig. 1, the three-phase voltage signals, $v_{abc}$, are sensed in time domain at the MG to measure the THD, named as $THD_{abc}$, obtain an estimate of the voltage amplitudes, defined as $\tilde{A}_{abc}$, and the zero-sequence components which defined as $V_{abc0}$, for fault detection and identification.

The THD is computed for each phase of the grid, $THD_{abc}$, according to the method reported in [15] which is obtained according the standard definition of [16, 17] that uses the square root of the sum of the squared harmonic components of a given signal, divided by the fundamental component:

$$\text{THD} = \frac{\sqrt{\sum_h |A_h|^2}}{A_1} \cdot 100, \quad (1)$$

where $h$ is the harmonic order and $A_h$ is the amplitude of the $h$-th harmonic component, with $h \neq 1$, and $A_1$ is the amplitude of the fundamental component.

Fig. 2. Depicts the block diagram of the THD method, composed by few blocks: second order generalized integrator (SOGI) grid monitoring system [18], a LPF and few math operations. The SOGI is used to provide an estimate of $A_1$ and the rest of harmonic components contained in the voltage signal, named as $e(t)$.

The zero-sequence voltages are used to identify between 2PH and 2PH-G, as both have the same conditions at the fault starting. The zero-sequence voltages are calculated as follows:

$$V_{abc0} = \frac{1}{3}(v_a + v_b + v_c) \quad (2)$$

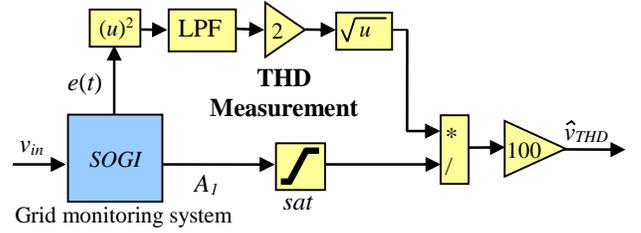

Fig. 2. THD measurement method block diagram

## B. Detection Stage

The fault detection is made, at the middle of Fig. 1, by using a threshold comparison with the $THD_{abc}$ voltages that had been measured in the previous stage. For the fault identification $\tilde{A}_{abc}$ and $V_{abc0}$ are also used. The algorithm is tested in a 0.42kV MG. The IEEE standard 519-2014 provides recommended values to limit the voltage harmonic distortion to 5% [19]. And according to the technical requirements in the Spanish grid code for reliable energy integration, the acceptable grid voltage drop range at the same level is set to 7.5 % [20]. Therefore, in this paper two thresholds are defined to help in fault detection and identification:

- $α$ to detect the fault when the $THD_{abc}$ surpass a 5%.
- $Δv$ to detect the fault when the estimate of the voltage amplitude $\tilde{A}_{abc}$ drops more than 7.5%.

## C. Decision Stage

In this stage, at the lower part of Fig. 1, the detection is done based on the behavior of $\tilde{A}_{abc}$, $THD_{abc}$, and $V_{abc0}$ as follows, and depending on the fault case. The faults had been classified into eleven categories numbered from 0 to 10 as shown in Table I.

TABLE I. FAULT TYPES CLASSIFICATION

| Fault Case | | Digital output of the algorithm |
|---|---|---|
| No fault | | 0 |
| Single phase-to-ground fault | AG | 1 |
| | BG | 2 |
| | CG | 3 |
| Phase to phase fault | AB | 4 |
| | BC | 5 |
| | CA | 6 |
| Phase-to-phase-to-ground fault | ABG | 7 |
| | BCG | 8 |
| | CAG | 9 |
| Three phase fault | ABG | 10 |

### 1) Symmetrical faults

These faults affect the three phases equally. The faults can happen between the three phases to ground (3PH-G) or between the three phases (3PH). In both cases, there are an abrupt

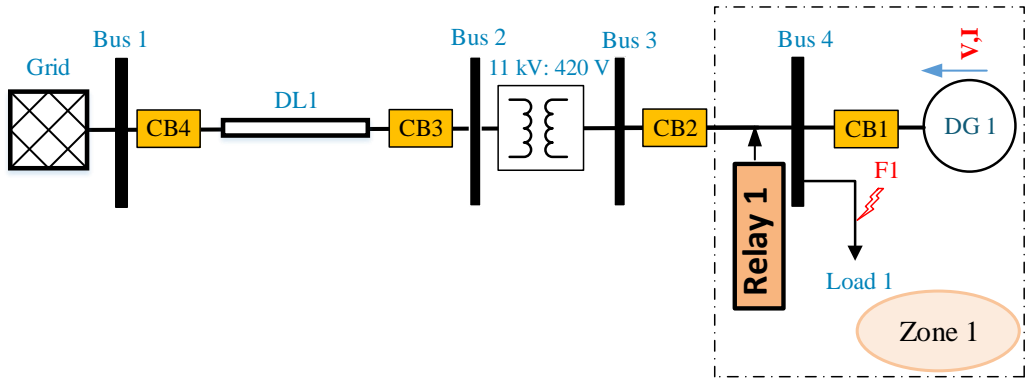

Fig. 3. Single line diagram of the studied electrical Network

increase in $THD_{abc}$ and, at the same time, a sudden decrease in $\tilde{A}_{abc}$ to 0 pu.

*2) Unsymmetrical faults*

Unsymmetrical faults cause an imbalance between the phases, which regarding the ground can be classified into:

*a) Phase-to-ground faults*: These faults can occur in two of the phases to ground (2PH-G), or in only one of the phases to ground (1PH-G). In this case, there is an abrupt increase in the measured THDs and, at the same time, a sudden drop in the estimated amplitudes to 0 pu that happen at the phases affected by a fault.

*b) Phase-to-phase faults:* These faults can happen between two of the phases (2PH), between a-to-b, a-to-c, or b-to-c phases. Then, an abrupt increase in the THD and a sudden drop in the estimated amplitudes of two of the grid phases is produced. However, in this cases, unlike the other grounded fault cases (i.e 2PH-G), and due to the absence of the ground connection and the low impendence between the faulted phases the estimated amplitude voltages go down just to 0.5 pu. As there are no zero-sequence sources in the 2PH faults [21], then if $V_{abc0} = 0$ the fault consists in a 2PH and to 2PH-G if $V_{abc0} \neq 0$.

## III. CASE STUDY

The electric system used for testing the behavior of the algorithm is shown in Fig. 3 and the parameters are listed in Table II. The system is composed by a 11kV grid with a Distribution Line (DL1), passing through a step-down transformer connected to a MG. The DL has the breakers (CB3 and CB4) that allows the disconnection in a fault event. A Distributed Generator (DG1) and local load (Load1) are connected to form a MG in Zone 1, that has its own relay and breakers (Relay1, CB1 and CB2). The algorithm is defined inside the relay for fault detection.

TABLE II. SYSTEM PARAMETERS

| Main Grid | MV/LV Transformer (Dyn11) | Distribution line DL1 | | DG1 Rating | Load1 Rating |
|---|---|---|---|---|---|
| Rated voltage 11kV | Rated power 400kVA | R | 0.16Ω/km | 77kVA | 480kW |
| | Rated Voltage 11/0.42kV | L | 0.109H/km | | |
| | | C | 0.31μF/km | | |
| | | Length 4km | | | |

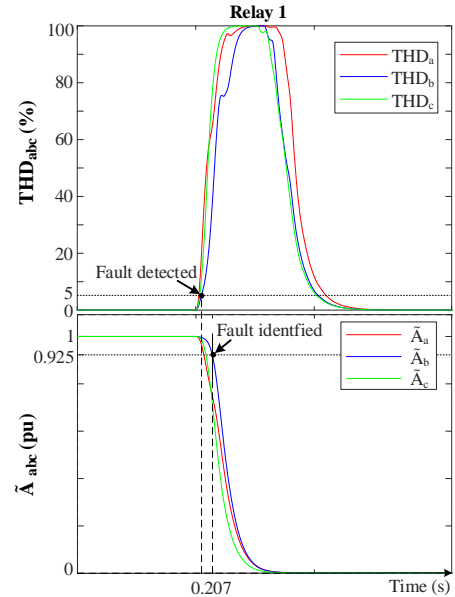

Fig. 4. Detection algorithm behavior during the 3PH-G fault at F1. Upper: THDs. Lower : estimated amplitudes $\tilde{A}$.

## IV. SIMULATION RESULTS

Simulations had been performed using MATLAB/Simulink at steady state to validate the performance of the detection algorithm under fault events. In this section, the fault cases are carried out in F1 location of Fig. 3 at 0.2s.

### A. Three phase fault (3PH-G)

Fig. 4 shows the detection algorithm behavior during the fault. In the healthy condition, when the system operates normally before the faults, the voltages have no harmonics, so the waveforms are sinusoidal. The measured THD is 0 and the estimated amplitude voltages are 1 pu. In this condition, the algorithm is waiting for an event of fault, therefore no detection action is performed.

At 0.2s, the $THD_{abc}$ increase abruptly, and when the condition $THD_{abc} > 5\%$ is meet the fault is detected. At the same time, $\tilde{A}_{abc}$ drops towards 0 pu and when $\tilde{A}_{abc} < 0.925$ pu, at this moment and due to the achievement of the two conditions, the fault is identified. Notice that $THD_{abc}$ have a behaviour that creates a peak and after a short time exponentially decays to zero. The detection process has been measured and it takes 7ms as shown in Fig. 5.

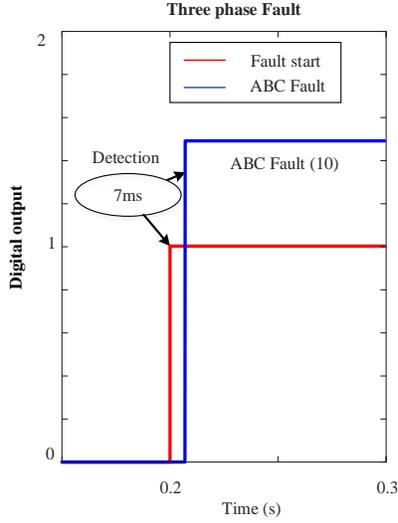

Fig. 5 Digital outputs of the detection algorithm in case of 3PH fault.

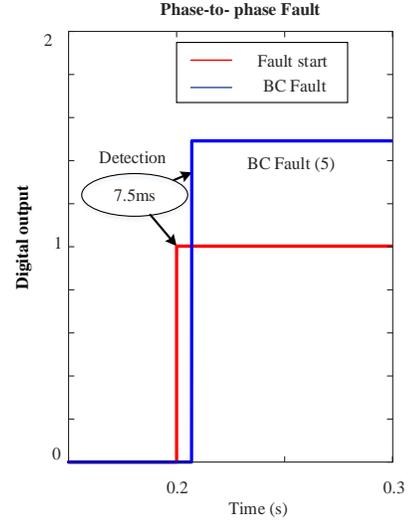

Fig. 7 Digital outputs of the detection algorithm in case of 3PH fault.

*B. Phase-to-phase fault (2PH)*

A fault between phases *b* and *c*, BC-fault, is considered in this case. Fig. 6 depicts the algorithm behavior during the fault. As in the previous case, at 0.2s $THD_{bc}$ increase abruptly, which makes the fault to be detected when $THD_{bc} > 5\%$, while $THD_a < 5\%$. Meanwhile, $\tilde{A}_{bc}$ drops towards 0.5 pu, due to the absence of the ground connection and to the impedance between the phases ($Z_f = 1m\Omega$), while $\tilde{A}_a$ remains unaffected (1 pu). Then, when $\tilde{A}_{bc} < 0.925$ pu and $V_{abc0}$ is checked to be zero, the fault is identified.

Similar to the previous case, the $THD_{bc}$ peak behaviour exponentially decays to zero after a short time. The detection process has been measured and it consists in 7.5ms as shown in Fig. 7.

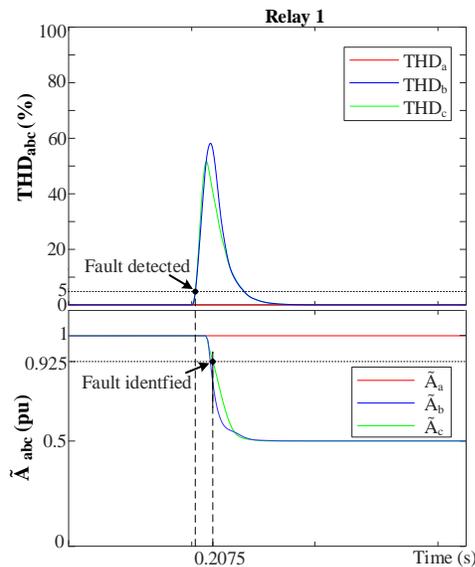

Fig. 6. Detection algorithm behavior during the 2PH fault at F1. Upper: THDs. Lower: estimated amplitudes $\tilde{A}$.

## V. CONCLUSION

This paper presents a fault detection algorithm for MGs that is based on the total harmonic behavior of the grid voltages. The THD levels of the grid voltages, the estimated amplitude voltages, and the zero-sequence components have been used to design the algorithm. Each phase in the system has its own measurement block to provide the necessary data to be mentored in the algorithm.

The MATLAB/Simulink simulations results show that the algorithm has the capability to detect and identify different types of faults that might occur in the electric system in the least possible time. The detection time in all the fault cases is less than 10 ms. The method used for obtaining the THD supposes a low computational burden for being implemented in a digital signal processor.